\documentclass[floatfix,twocolumn,showpacs,aps,tightenlines,superscriptaddress]{revtex4-1}
\usepackage{graphicx}
\usepackage{color}
\usepackage{psfrag}
\usepackage{dcolumn}
\usepackage{bm}

\usepackage{amssymb}
\usepackage{amsmath}
\usepackage{amsfonts}
\usepackage{longtable}
\usepackage{xspace}
\usepackage{verbatim}

\newcommand{\beq}{\begin{equation}}
\newcommand{\eeq}{\end{equation}}

\newcommand{\be}{\begin{equation}}
\newcommand{\ee}{\end{equation}}
\newcommand{\bea}{\begin{eqnarray}}
\newcommand{\eea}{\end{eqnarray}}
\newcommand{\bes}{\begin{subequations}}
\newcommand{\ees}{\end{subequations}}

\newcommand{\MPA}{{moving punctures approach}\xspace}
\newcommand{\hispid}{{\sc HiSpID}\xspace}

\usepackage{bm}

\begin{document}

\title{Evolutions of nearly maximally spinning black hole binaries using the moving
puncture approach}

\author{Yosef Zlochower} 
\author{James Healy} 
\author{Carlos O. Lousto}
\affiliation{Center for Computational Relativity and Gravitation,
School of Mathematical Sciences,
Rochester Institute of Technology, 85 Lomb Memorial Drive, Rochester,
 New York 14623}
\author{Ian Ruchlin} 
\affiliation{Department of Mathematics, West Virginia University,
Morgantown, West Virginia 26506, USA}

\date{\today}

\begin{abstract}
  We demonstrate that numerical relativity codes based on the {\it
  moving punctures} formalism are capable of evolving nearly maximally
  spinning black hole binaries. We compare a new evolution of an
  equal-mass, aligned-spin binary with dimensionless spin $\chi=0.99$
  using puncture-based data with recent simulations of the SXS
  Collaboration. We find that the overlap of our new waveform with the
  published results of the SXS Collaboration is larger than 0.999.
  To generate our new waveform, we use the recently introduced
  \hispid puncture data, the CCZ4 evolution system, and a modified
  lapse condition that helps keep the horizon radii reasonably large.
\end{abstract}

 \maketitle

\section{Introduction}\label{sec:intro}

Since the breakthroughs in numerical relativity of
2005~\cite{Pretorius:2005gq, Campanelli:2005dd, Baker:2005vv} it is
possible to accurately simulate
moderate-mass-ratio and moderate-spin black-hole
binaries. State of the art numerical relativity
codes now routinely evolve
binaries with mass ratios as small as $q\lesssim
1/20$~\cite{Gonzalez:2008bi, Lousto:2010qx, Lousto:2010ut,
Sperhake:2011ik, Chu:2015kft, Jani:2016wkt}, and are pushing towards
much smaller mass ratios. Indeed, there have been several explorations
of
$q=1/100$ binaries~\cite{Lousto:2010ut, Sperhake:2011ik}.

However, when it comes to highly-spinning binaries,
prior to the work of~\cite{Lovelace:2008tw} of the SXS
Collaboration \footnote{{\tt https://www.black-holes.org}}, it
was not even possible to construct initial data for binaries with
spins larger than~$\sim 0.93$~\cite{Cook:1989fb}. This limitation was
due to the use of conformally flat initial data. Conformal flatness is
a convenient assumption because the Einstein constraint system takes on
particularly simple forms. Indeed, using the puncture approach, the
momentum constraints can be solved exactly using the Bowen-York
ansatz~\cite{Bowen:1980yu}. There were several attempts to increase
the spin of black hole, while still preserving conformal
flatness~\cite{Dain:2002ee, Lousto:2012es}, but these introduced
negligible improvements. Lovelace {\it et al.}~\cite{Lovelace:2008tw}
were able to overcome these limitations by choosing the initial data
to be a superposition of conformally Kerr black holes in the
Kerr-Schild gauge. Using these
new data, they were soon able to evolve binaries with spins as large
as
0.97~\cite{Lovelace:2011nu} and, later, spins as high as
0.994~\cite{Scheel:2014ina}.

While spins of $0.92$ may seem reasonably close to 1, the scale is
misleading. The amount of rotational energy in a black hole with spin
0.9 is only 52\% of the maximum. Furthermore, particle limit and
perturbative calculations show even more extreme differences between
spins of 1 and spins only slightly smaller. For example, Yang {\it et
al.}~\cite{Yang:2014tla} studied an analog to turbulence in black-hole perturbation
theory. For spins close to 1, there is an inverse energy cascade from
higher azimuthal ($m$) modes to lower ones for $\ell$ modes that obey
 $\epsilon = |1 - \chi|  \lesssim \ell^{-2}$. This gives
hints that a more useful measure of the spin is actually
$1/\epsilon$. Similarly, both the analysis of Kerr
geodesics~\cite{Schnittman:2014zsa,
Berti:2014lva} and  particle-limit calculations of
recoils~\cite{Hirata:2010xn, vandeMeent:2014raa}, indicate that the
dynamics of nearly extremal-spin black
holes cannot be elucidated with any degree of certainty using
lower spin simulations.

Another area of interest is the use of numerical relativity waveforms
in the detection and parameter estimation of gravitational wave signals
as observed by LIGO and other detectors \cite{Abbott:2016apu}.
This important region of
parameter space with highly spinning binaries is currently poorly covered
and will benefit from new and accurate simulations. 

Recently, we introduced a version of
highly-spinning initial data, also based on the superposition of two
Kerr black holes~\cite{Ruchlin:2014zva, Healy:2015mla}, but this time
in a puncture gauge. The main differences between the two approaches
is how easily the latter can be incorporated into moving-punctures
codes. In Ref.~\cite{Ruchlin:2014zva}, we were able to evolve an
equal-mass binary with aligned spins, and spin magnitudes of
$\chi=0.95$, using this new data and compare with the results of the
Lovelace {\it et al.}.

Prior to our work, Hannam {\it et al.}~\cite{Hannam:2006zt} considered
the case of non-boosted, highly spinning black holes. Similar to what
we see here, they found that removing the conformally flat ansatz
greatly reduces the amount of unphysical radiation.

In this paper, we show the results of a simulation of an equal-mass
binary with aligned spins of $\chi=0.99$. We compare the 
$(\ell=2, m=2)$  and $(\ell=3, m=m2)$ modes of the waveform
with those previously published by the SXS
Collaboration in~\cite{Scheel:2014ina}.
This comparison  allows  us to assess the errors in
these waveforms and to gain confidence about reaching the
required accuracy for use in gravitational wave astronomy.

We use the following standard conventions throughout this paper.
In all cases, we use geometric units where $G=1$ and $c=1$. 
Latin letters ($i$, $j$, $\ldots$) represent spatial indices.
Spatial 3-metrics are denoted by $\gamma_{ij}$ and extrinsic
curvatures by $K_{ij}$. The trace-free part of the extrinsic curvature
is denoted by $A_{ij}$. A tilde indicates a conformally related
quantity. Thus $\gamma_{ij} = \psi^4 \tilde \gamma_{ij}$ and $A_{ij} =
\psi^{-2} \tilde A_{ij}$, where $\psi$ is some conformal factor. We
denote the covariant derivative associated with $\gamma_{ij}$ by $D_i$
and the covariant derivative associated with $\tilde \gamma_{ij}$ by
$\tilde D_i$. A lapse function is denoted by $\alpha$, while a shift
vector by $\beta^i$.

This paper is organized as follows. In Sec.~\ref{sec:ID}, we provide a
brief overview of how the initial data are constructed. In
Sec.~\ref{sec:evolution} we describe the numerical techniques used to
evolve these data. In Sec.~\ref{sec:results}, we compare the new
\hispid waveform with a similar SXS waveform. In
Sec.~\ref{sec:diagnostic}, we analyze the various diagnostics to
determine the accuracy of the simulation. Finally, in
Sec.~\ref{sec:discussion}, we discuss our results.

\section{Numerical Techniques}\label{sec:techniques}

\subsection{Initial Data}\label{sec:ID}
We construct initial data for a black-hole binary with individual
spins $\chi_{1,2} = 0.99$ using the \hispid 
code~\cite{Ruchlin:2014zva, Healy:2015mla}. The \hispid code solves
the four Einstein constraint equations using the   conformal
transverse traceless
decomposition~\cite{York99, Cook:2000vr, Pfeiffer:2002iy,
AlcubierreBook2008}. In this approach, the spatial metric
$\gamma_{ij}$ and extrinsic curvature $K_{ij}$ are given by
\begin{eqnarray}
  \gamma_{ij} = \psi^4 \tilde \gamma_{ij},\\
  K_{ij} = \psi^{-2}\tilde A_{ij} + \frac{1}{3} K \gamma_{ij},\\
  \tilde A_{ij} = \tilde M_{ij} + (\tilde{\mathbb{L}} b)_{i j},
\end{eqnarray}
where the conformal metric $\tilde \gamma_{ij}$, the trace of the
extrinsic curvature $K$, and the trace-free tensor $\tilde M_{ij}$ are
free data. The Einstein constraints then become a set of four coupled
elliptical equations for the scalar field $u = \psi - \psi_0$ and components of the spatial vector
$b^i$ ($\psi_0$ is a singular function specified analytically).
The resulting elliptical equations are solved using an
extension to the \textsc{TwoPunctures}~\cite{Ansorg:2004ds} thorn.

The free data are chosen by superimposing two boosted Kerr black
holes, as described in more detail in~\cite{Ruchlin:2014zva}.
The superposition has the form
\begin{eqnarray}
    \tilde \gamma_{ij} = \tilde \gamma^{(+)}_{ij} +
    \tilde \gamma^{(-)}_{ij} - \delta_{ij},\\
    K = K^{(+)} + K^{(-)},\\
    M_{ij} = \left[\tilde A_{ij}^{(+)} + \tilde
    A_{ij}^{(-)}\right]^{\bf TF},\\
    \psi_0 = \psi_{(+)} + \psi_{(-)} - 1,
\end{eqnarray}
where $(+)$ and $(-)$ refer to the two black holes, $\tilde
\gamma^{(\pm)}_{ij}$ and $\tilde A_{ij}$ are the conformal metric and
trace-free extrinsic curvatures for a boosted and rotated Kerr black hole, $K^{(\pm)}$ is the
mean curvature, and the conformal factor $\psi_{(\pm)}$ is chosen such
that $\psi_{(\pm)} = \sqrt[12]{{\rm det}(\gamma_{ij}^{(\pm)})}$ (where
$\gamma_{ij}^{(\pm)}$ is the physical metric from a boosted and
rotated Kerr black
hole).

To get $\tilde \gamma_{ij}^{(\pm)}$, etc.,  we
start with Kerr black holes in quasi-isotropic (QI) coordinates and perform
a fisheye (FE) radial coordinate transformation (where $r_{\rm QI}=0$ is the
location of the puncture),
\begin{equation}
  r_{\rm QI} = r_{\rm FE} [1-A_{\rm FE} \exp(-r_{\rm FE}^2/{s_{\rm FE}}^2)],
\end{equation}
where $r_{\rm FE}$ is the fisheye radial coordinate, $r_{\rm QI}$ is
the original QI radial coordinate, and  $A_{\rm FE}$ and $s_{\rm FE}$ are parameters. These coordinates have the property
that at large $r_{\rm FE}$, $r_{\rm QI} \approx r_{\rm FE}$, and
at small $r_{\rm FE}$, $dr_{\rm QI} = (1-A) dr_{\rm FE}$ (i.e.,
$dr_{\rm QI} < dr_{\rm FE}$).
The FE transformation is needed because it expands the horizon size
from $r_h\approx 0.035$ to $r_h\approx0.5$. We then transform the
metric to Cartesian-like coordinates of the form
$x=r\sin\theta\cos\phi$, $y=r\sin\theta\sin\phi$, $z=r\cos\theta$,
where $r = r_{\rm FE}$.
We then perform a Lorentz-like boost on this metric and, in the
case of nonaligned spins, a rotation. The resulting 4-metric
is then decomposed into a spatial metric $\gamma_{ij}$ and extrinsic
curvature $K_{\ij}$.

We use the
$g$ attenuation described in~\cite{Ruchlin:2014zva} to modify both the
metric and elliptical equations inside the horizons.
We briefly summarize the procedure here.
The modified Hamiltonian and momentum constraint equations for
the correction functions $u$ and $b^i$ are
\begin{subequations}
  \label{eq:constraints}
  \begin{align}
    \tilde{D}^2 u - g \frac{\psi \tilde{R}}{8} - g \frac{\psi^{5} K^2}{12}
    + g \frac{\tilde{A}_{i j} \tilde{A}^{i j}}{8 \psi^{7}} + g
    \tilde{D}^2\left(\psi_{(+)} + \psi_{(-)}\right) &= 0 \; ,
    \label{eq:hamiltonian} \\
    \tilde{\Delta}_{\mathbb{L}} b^i + g \tilde{D}_{j} \tilde{M}^{i
    j} - g \frac{2}{3} \psi^{6} \tilde{\gamma}^{i j} \tilde{D}_{j}
    K &= 0 \; ,\label{eq:momentum}
  \end{align}
\end{subequations}
where $\tilde{\Delta}_{\mathbb{L}} b^i \equiv \tilde{D}_{j}
(\tilde{\mathbb{L}} b)^{i j}$ is the vector Laplacian and $\tilde{R}$
is the scalar curvature associated with $\tilde{\gamma}_{i j}$, and
where the attenuation function $g$ takes the form
\begin{align*}
g &= g_{+}\times g_{-} \; ,\\
  g_{\pm} &= 
          \begin{cases} 
     1 & \mbox{if } r_\pm > r_{\rm max} \\
     0 & \mbox{if } r_\pm < r_{\rm min} \\
            {\cal G}(r_{\pm}) & \mbox{otherwise},
  \end{cases} \; ,\\
  {\cal G}(r_\pm) &= \frac{1}{2}\left[1+ \tanh\left(\tan\left[ \frac{\pi}{2}
  \left(-1 + 2 \frac{r_{\pm}-r_{\rm min}}{r_{\rm max} -
  r_{\rm min}}\right)\right]\right)\right],
\end{align*}
$r_{\pm}$ is the coordinate distance to puncture $(+)$ or
$(-)$,
and the parameters $r_{\rm min} < r_{\rm max}$ are chosen
to be within the horizon. In addition, we attenuate the
background metric itself when calculating the
$\tilde D^2 u$ and $\tilde{\Delta}_{\mathbb{L}} b^i$. To do this,
we take
\begin{eqnarray}
  \tilde \gamma_{ij}  \to \delta_{ij} + g (\tilde \gamma_{ij} -
  \delta_{ij}),\\
  \tilde \Gamma^{k}_{\,ij} \to g \tilde \Gamma^{k}_{\,ij}.
\end{eqnarray}
Note that the modified $\tilde \Gamma^{k}_{\,ij}$ is not consistent
with the modified $\tilde \gamma_{ij}$. There is no advantage to
making them consistent because the constraints will be violated in the
attenuation zone regardless. By modifying the metric in this way, we
can ensure that the elliptical system has exactly the form of the flat
space Poisson system in the vicinity of the punctures.

Finally, far from the holes, we attenuate  $\tilde \gamma_{ij}$, $K$,
and $\psi_0$. This is achieved by consistently changing the metric
fields
and their derivatives so that
\begin{eqnarray}
  \tilde \gamma_{ij}^{(\pm)} \to f(r_\pm) (\tilde \gamma_{ij}^{(\pm)} - \delta_{ij}) +
  \delta_{ij},\\ 
  K^{(\pm)} \to f(r_\pm) K^{(\pm)},\\
\left(\psi_{(\pm)}-1\right) \to f(r_\pm) \left(\psi_{(\pm)}-1\right),
\end{eqnarray}
where $f(r) = \exp(-r^4/s_{\rm far}^4)$ and $r_{\pm}$ is the
coordinate distance to puncture $(+)$ or $(-)$.
                                                                    
For compatibility with the original {\sc TwoPunctures} code, we chose
to set up \hispid so that the parameters of the binary are specified
in terms of momenta and spins of the two holes. However, unlike for
Bowen-York data, the values specified are only approximate, as the
solution vector $b^i$ can modify both of these. In practice, we find
that the spins are modified by only a trivial amount while
orbital angular momentum is reduced significantly.
We compensate for this by choosing larger momentum parameters
than those predicted by simple quasicircular conditions would
imply~\cite{Healy:2017zqj}. All parameters for the $\chi=0.99$ run are
given in Table~\ref{tab:id}.
The quantity $r_{H}$ in the table is the 
polar coordinate radius (which is the smallest radius on each
horizon). As
this is gauge dependent, it can change arbitrarily during the
evolution. However, large changes are generally undesirable. The size
of $r_H$ is also directly related to the number of refinement levels
required, and therefore to the computational cost. An ideal gauge
would have $r_H$ settle to a moderate value and remain
there. The initial size of the horizon is chosen to be large in order
to speed up the convergence of the initial data solver (this is due to
the scale set by the $g$ attenuation discussed above). However, the
gauge conditions we use quickly drive $r_{H}$ towards smaller values.
We note that in quasi-isotropic coordinates, the coordinate radius of
a maximally spinning black hole is zero.

\begin{table}
  \caption{Initial data parameters for a $\chi=0.99$ highly spinning
  binary. The two spins are given by $\vec S_{1,2} = (0,0,S)$ and the
two momenta are $P_{1,2} = \pm (0, P,0)$.  The parameter ${\cal M}$ is
the mass of the two black holes. Unlike for Bowen-York data, the
momenta and spins cannot be specified exactly. However, the mass
${\cal M}$ is very close to the measured horizon mass $m_{H}$.
Quantities denoted by ``init'' were measured at
$t=0$, while quantities denoted by ``equi'' are averaged over the
several orbits. $A_{\rm FE}$, $s_{\rm FE}$, $r_{\rm min}$, $r_{\rm
max}$, and $s_{\rm far}$ are attenuation parameters. $m_H$, $S$, $\chi$ are masses, spin angular momenta,
and dimensionless spins, respectively, of the two black holes. $M_{\rm
rem}$ and $\chi_{\rm rem}$ are the remnant mass and dimensionless
spin. The quantity $r_{H}$ is the polar coordinate radius of the
horizons.
Finally, $M_{\rm ADM}$ and $J_{\rm ADM}$ are the ADM masses and
spins.
}\label{tab:id}
  \begin{ruledtabular}
    \begin{tabular}{llllll}
      ${\cal M}/M = 0.505570 $ &  $P/M =  0.09675$ \\
      $S/M^2 = 0.253045$ & $A_{\rm FE}= 0.99$ \\
      ${s_{\rm FE}} =  1.7$ & $r_{\rm min}= 0.01$  \\
      $r_{\rm max}= 0.40$  & $s_{\rm far} = 10$\\
      \\
      $J_{\rm ADM}/M^2 =
      1.42621$ &
      $M_{\rm ADM}/M = 0.99998$\\
      $m_{H\ \rm init}/M = 0.50555$ & $m_{H\ \rm equi}/M =
      0.5072\pm0.0004$ \\
    $S_{\rm init}/M^2 = 0.2529$ & $S_{\rm equi}/M^2 = 0.2547\pm0.0004$ \\
    $\chi_{\rm init} = 0.9897$ & $\chi_{\rm equi} = 0.9903\pm0.0002$ \\
      $r_{H\ \rm init}/M = 0.44$ & $r_{H\ \rm equi}/M = 0.082\pm0.001$\\
      \\
      $M_{\rm rem}/M  = 0.898\pm0.001$ & $\chi_{\rm rem} =
      0.949\pm0.001$\\
    \end{tabular}
  \end{ruledtabular}
\end{table}

\subsection{Evolution}\label{sec:evolution}

We evolve black hole binary initial data sets using the 
{\sc LazEv}~\cite{Zlochower:2005bj} implementation of the \MPA 
for the conformal and covariant formulation of the Z4 (CCZ4) system
(Ref.~\cite{Alic:2011gg}) which includes stronger damping of
the constraint violations than the standard BSSNOK~\cite{Nakamura87, Shibata95,
Baumgarte99} system.
For the run presented here, we use
centered, eighth-order accurate finite differencing in
space~\cite{Lousto:2007rj} and a fourth-order Runge-Kutta time
integrator. Our code
uses the {\sc Cactus}/{\sc EinsteinToolkit}~\cite{cactus_web,
einsteintoolkit} infrastructure.  We use the {\sc Carpet} mesh 
refinement driver to provide a ``moving boxes'' style of mesh refinement
\cite{Schnetter-etal-03b}.  Fifth-order Kreiss-Oliger dissipation is added to
evolved variables with dissipation coefficient $\epsilon=0.1$.
For the CCZ4 damping parameters, we chose
$\kappa_1 = 0.2$, $\kappa_2=0$, and $\kappa_3=0$
(see~\cite{Alic:2011gg}), but found that these had to be modified
during the evolution.

We locate the apparent horizons using the {\sc AHFinderDirect}
code~\cite{Thornburg2003:AH-finding} and measure the horizon spins
using the isolated horizon (IH) algorithm~\cite{Dreyer02a}.
We calculate the radiation scalar $\psi_4$ using the Antenna
thorn~\cite{Campanelli:2005ia, Baker:2001sf}.
We then extrapolate the waveform to
an infinite observer location using
the perturbative formulas given in Ref.~\cite{Nakano:2015pta}.

For the gauge equations, we use~\cite{Alcubierre02a,
Campanelli:2005dd, vanMeter:2006vi}
\begin{subequations}
    \label{eq:gauge}
      \begin{align}
           (\partial_t - \beta^i \partial_i) \alpha &= - 2 \alpha^2 K \; , \\
            \partial_t \beta^a &= \frac{3}{4} \tilde{\Gamma}^a - \eta
        \beta^a \; .
          \end{align}
\end{subequations}
Note that the lapse is not evolved with the standard 1+log form. Here
we multiply the rhs of the lapse equation by an additional factor of
$\alpha$. This has the effect of increasing the equilibrium
(coordinate) size of the horizons. For the initial values of shift, we
chose $\beta^i(t=0) =0$, while for the initial values of the lapse, we
chose an ad-hoc function
$\alpha(t=0) = \tilde \psi^{-2}$, where
$\tilde \psi = 1 + {\cal M}/(2 r_1) + {\cal M}/(2 r_2)$ and
$r_i$ is the coordinate distance to BH $i$. For the function $\eta$,
we chose
\begin{equation}
  \eta(\vec r) = (\eta_c - \eta_o) \exp(-(r/\eta_s)^4) + \eta_o,
\end{equation}
where
$\eta_c = 2.0/M$, $\eta_s = 40.0M$, and $\eta_o = 0.25/M$. With this
choice, $\eta$ is small in the outer zones. As shown in
Ref.~\cite{Schnetter:2010cz}, the magnitude of $\eta$ limits how large
the timestep can be with $dt_{\rm max} \propto 1/\eta$. Since this
limit is independent of spatial resolution, it is only significant in
the very coarse outer zones where the standard Courant-Friedrichs-Lewy
condition would otherwise lead to a large value for $dt_{\rm max}$.

We evolved the $\chi=0.99$ data using 11 levels of refinement, with
the outermost grid extending to $400M$ with a gridspacing of $2.78M$.
The gridspacing on the finest grid was $h=M/368.64$. The total cost of
the simulation was 710 KSU.
\begin{figure}
  \includegraphics[width=0.9\columnwidth]{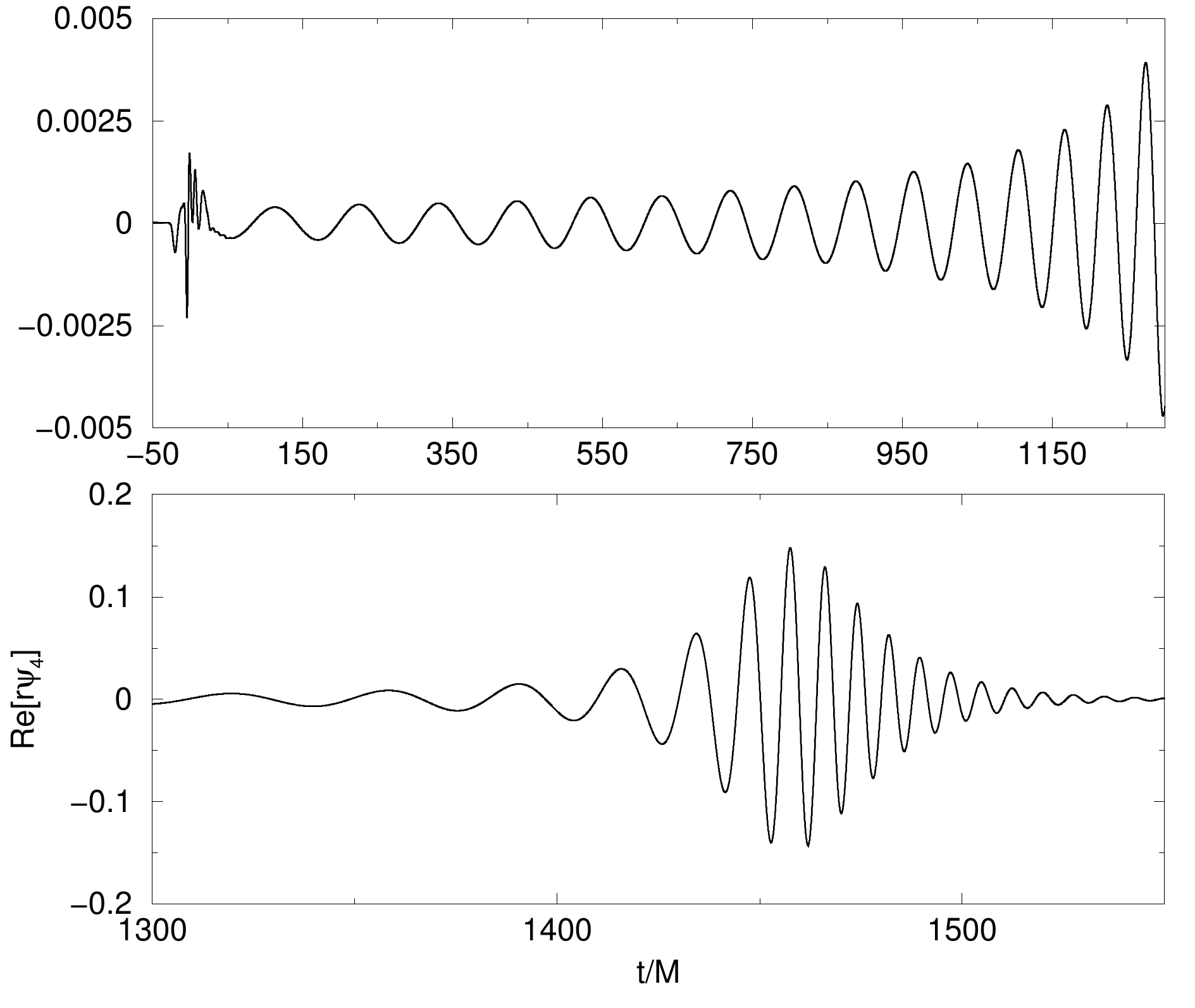}
  \caption{The \hispid waveform showing the amplitude of the initial
    data pulse compared to the physical waveform. Note how little the
  pulse contaminates the rest of the signal.}\label{fig:full_wave}
\end{figure}

One remarkable consequence of these superimposed Kerr data is how
small the initial pulse of unphysical radiation is. As first seen in
the nonboosted case by Hannam {\it et al.}~\cite{Hannam:2006zt}, the
initial pulse is roughly four times as large as the orbital signal at a
separation of $D=10M$. While this may sound quite large, for a
$\chi=0.9$ binary, the amount of unphysical radiation for a Bowen-York
binary is six times more, and it rapidly increases with spin. The full
waveform, including initial pulse, is shown in
Fig.~\ref{fig:full_wave}.

\section{Results}\label{sec:results}

\begin{widetext}
We performed a single simulation from a coordinate separation of $10M$
(proper separation of $12.2M$) through merger for an equal-mass binary
where both spins are aligned with the orbital angular momentum and
have dimensionless magnitudes of 0.99. We compare these with the
BBH:0177 waveform~\cite{Scheel:2014ina, SXS:catalog}. In order to
compare the \hispid and SXS waveform, we rescale the time coordinate
by the ratio of the final masses and then introduce a constant phase
and time translation to minimize the RMS difference between the two
waveforms. We note that the SXS waveform is longer by about $t=5000M$.
\begin{figure*}[h!]
    \includegraphics[width=0.51\textwidth]{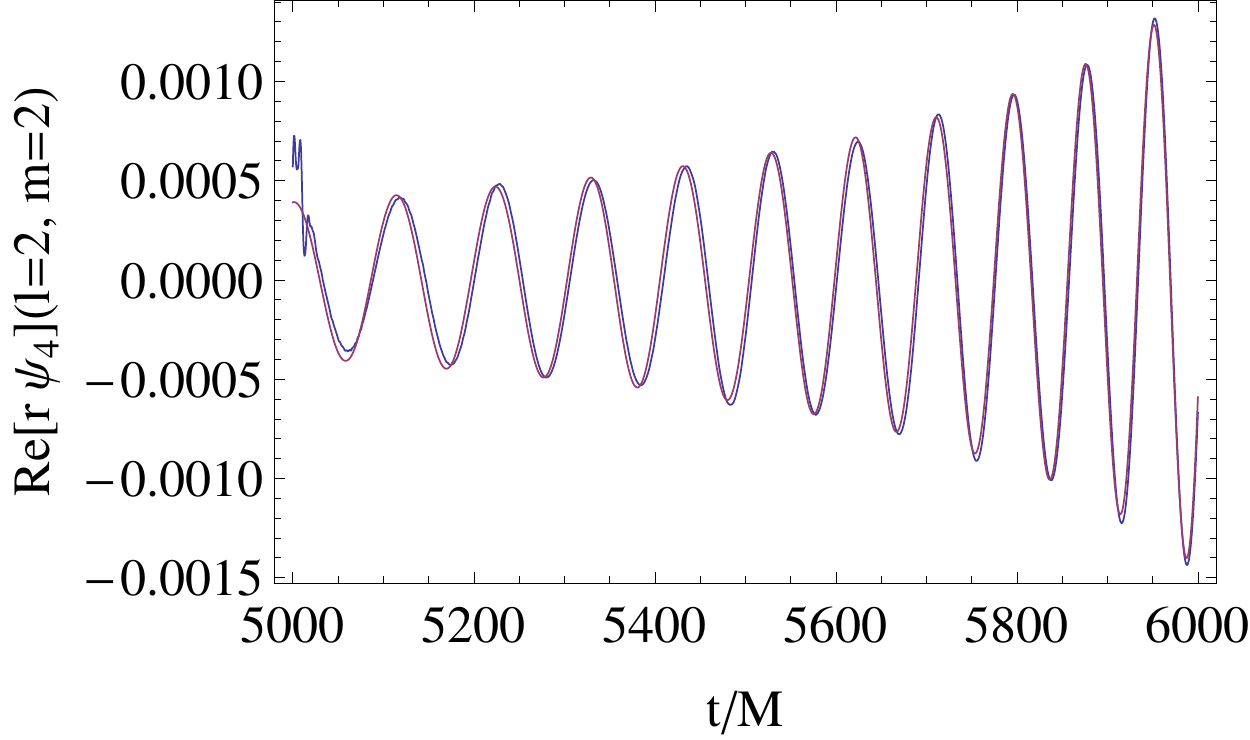}
  \includegraphics[width=0.48\textwidth]{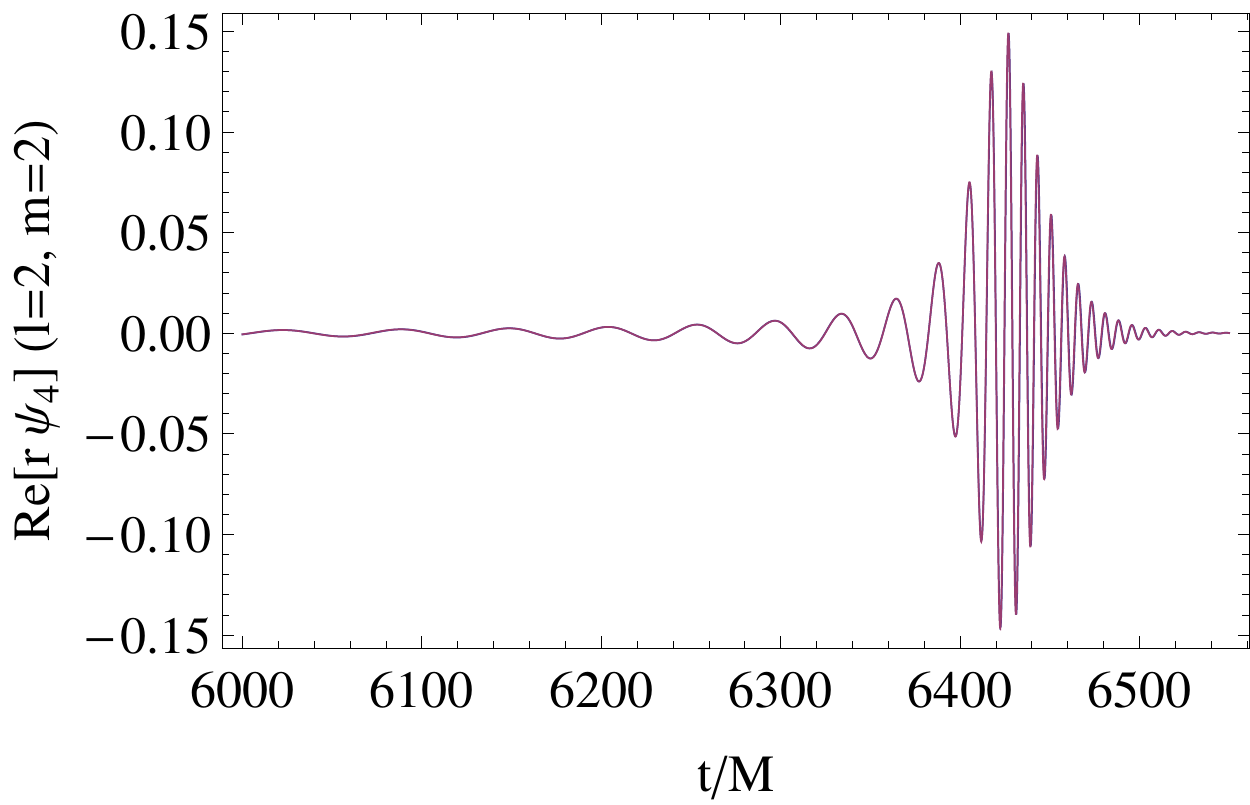}\\
  \includegraphics[width=0.48\textwidth]{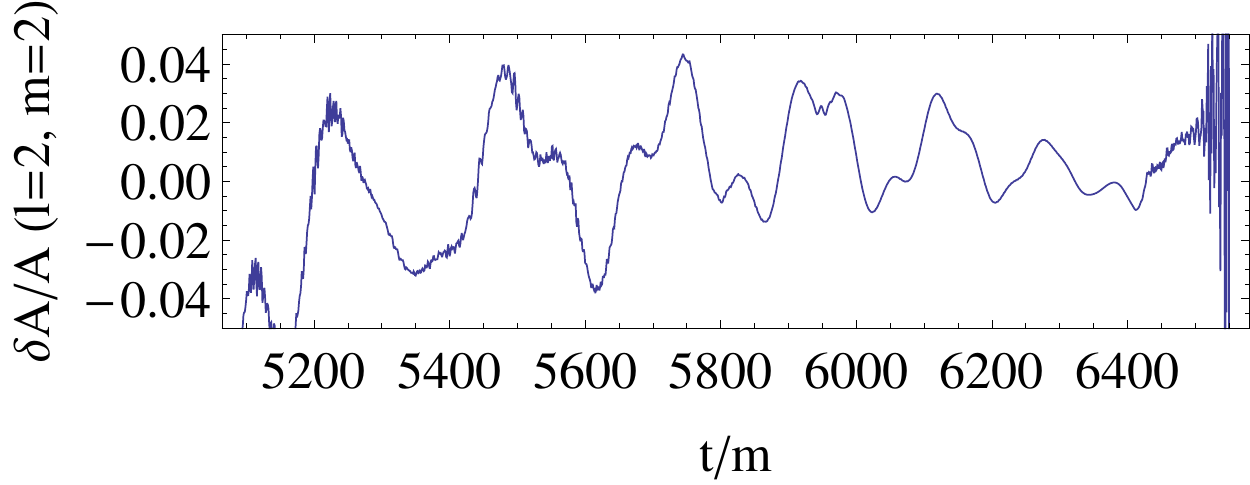}
  \includegraphics[width=0.48\textwidth]{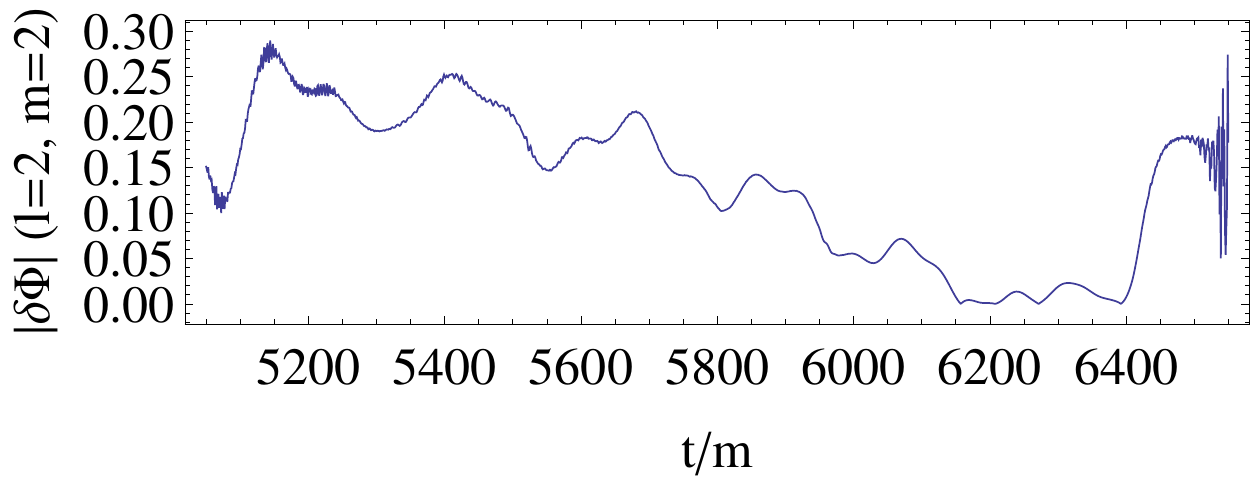}
  \caption{(Top left and top right) The new HiSpID simulation (blue) and the SXS simulation
    (red) of the $(\ell=2, m=2)$ mode of $\psi_4$ (real part). The HiSpID waveform was translated by
  $t\sim5000M$. (Bottom left) The difference in amplitude between the
HiSpID and
    SXS waveforms. (Bottom right) The difference in phase between the
  HiSpID
    and
  SXS waveforms. Note that the period of oscillations in $\delta A/A$ and
  $\delta \Phi$ is very close to the orbital timescale (see
  Fig.~\ref{fig:ecc}). This indicates that these oscillations are likely due
  to eccentricity.
}\label{fig:RIT_SXS_COMP_WAVE}
\end{figure*}
\end{widetext}

\begin{figure}
  \includegraphics[width=0.8\columnwidth]{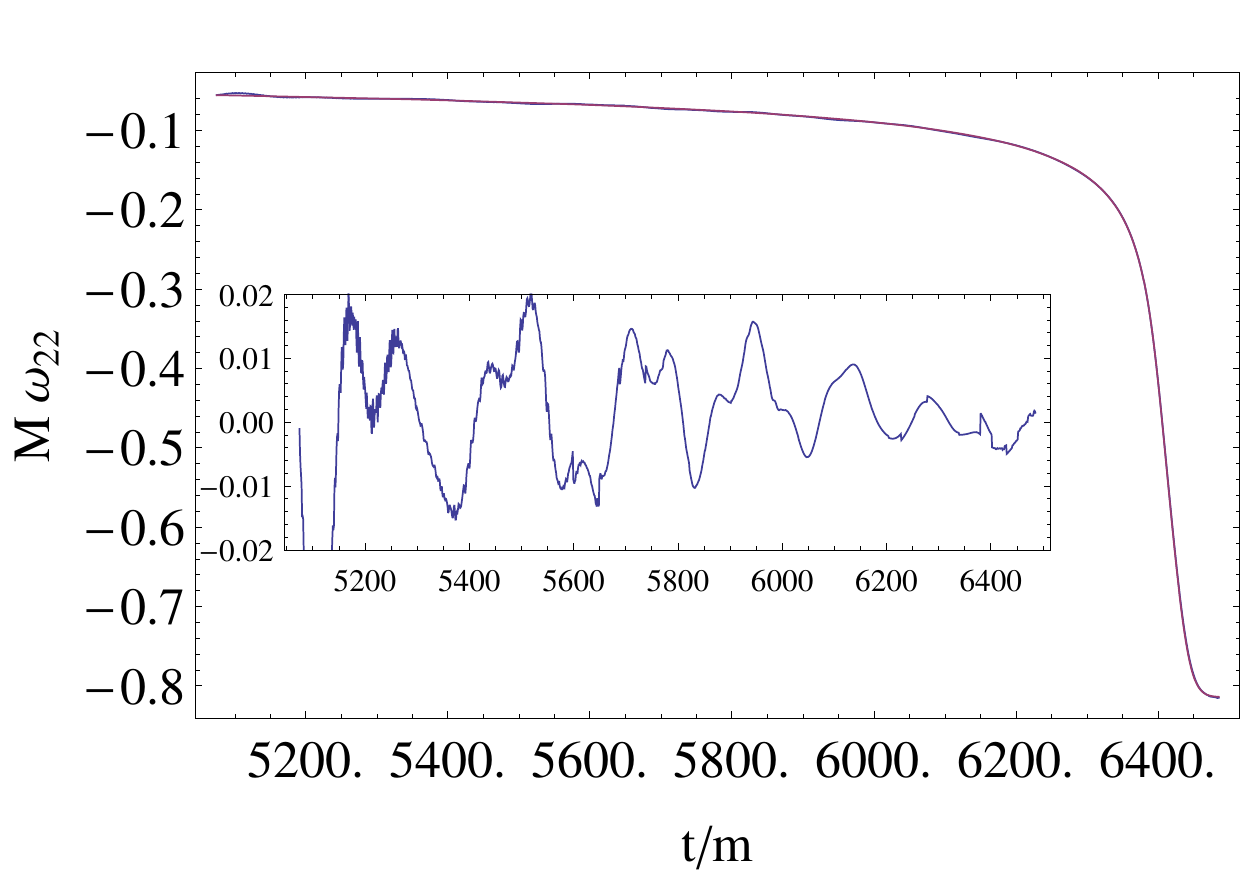}
  \caption{The frequency of the $(\ell=2,m=2)$ mode of $\psi_4$
    for the \hispid (blue) and SXS (red) waveforms. The inset shows
    the relative differences in frequency between the two waveforms.
  }\label{fig:RIT_SXS_COMP_FREQ}
\end{figure}

In Figs.~\ref{fig:RIT_SXS_COMP_WAVE} and \ref{fig:RIT_SXS_COMP_FREQ}, we directly compare the new
\hispid waveform with the corresponding SXS waveform. We translate the \hispid waveform to
maximize the overlap. Hence, the ``starting" time in the figures is
$t\sim5000M$. The overlap is defined by~\cite{Campanelli:2008nk}
\begin{equation}
  \underset{t_0}{\rm MAX} \frac{\left|\langle R(t),
  S(t+t_0)\rangle \right|}{\sqrt{\left|\langle R(t),
  R(t)\rangle \right|\ \left|\langle S(t+t_0),
    S(t+t_0)\rangle \right|}},
\end{equation}
where
\begin{equation}
  \langle a(t), b(t)\rangle = \int_0^{t_f} \bar a(t) b(t) dt,
  \end{equation}
  an overbar denotes complex conjugation,
and $t_0$ is chosen to maximize the result, while
$t=0$ corresponds to the time just after the initial pulse has
radiated away and $t_f$ to the last timestep in the \hispid
simulation. We find an overlap of $0.99975$ for the ($\ell=2, m=2)$
mode, which is quite good
considering that the \hispid waveform is eccentric ($e\sim0.01$), while
the SXS waveform is not. Note the phase agreement is within 0.25 rad
across the entire waveform and the amplitude agreement is better than
4\%. The agreement in frequency is even better, with a relative
difference of lass than 2\% across the entire waveform.

\begin{figure}[h!]
    \includegraphics[width=0.99\columnwidth]{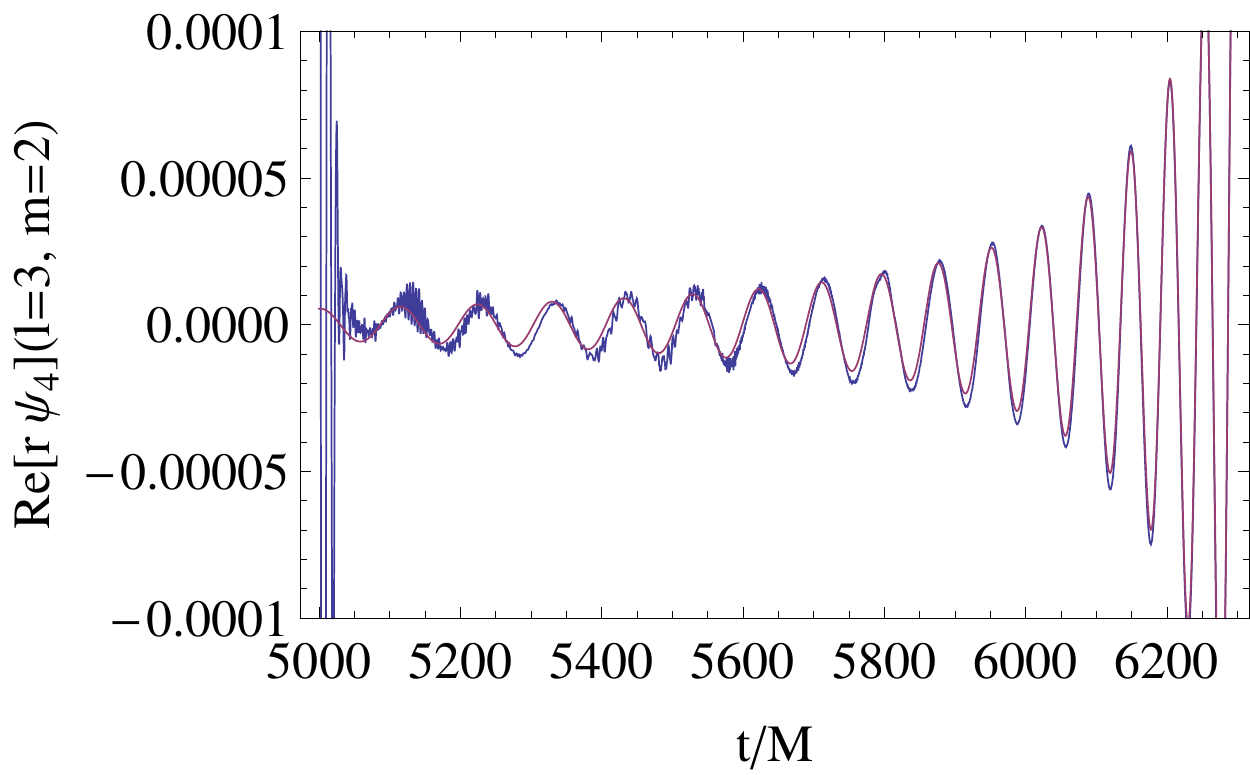}
  \includegraphics[width=0.99\columnwidth]{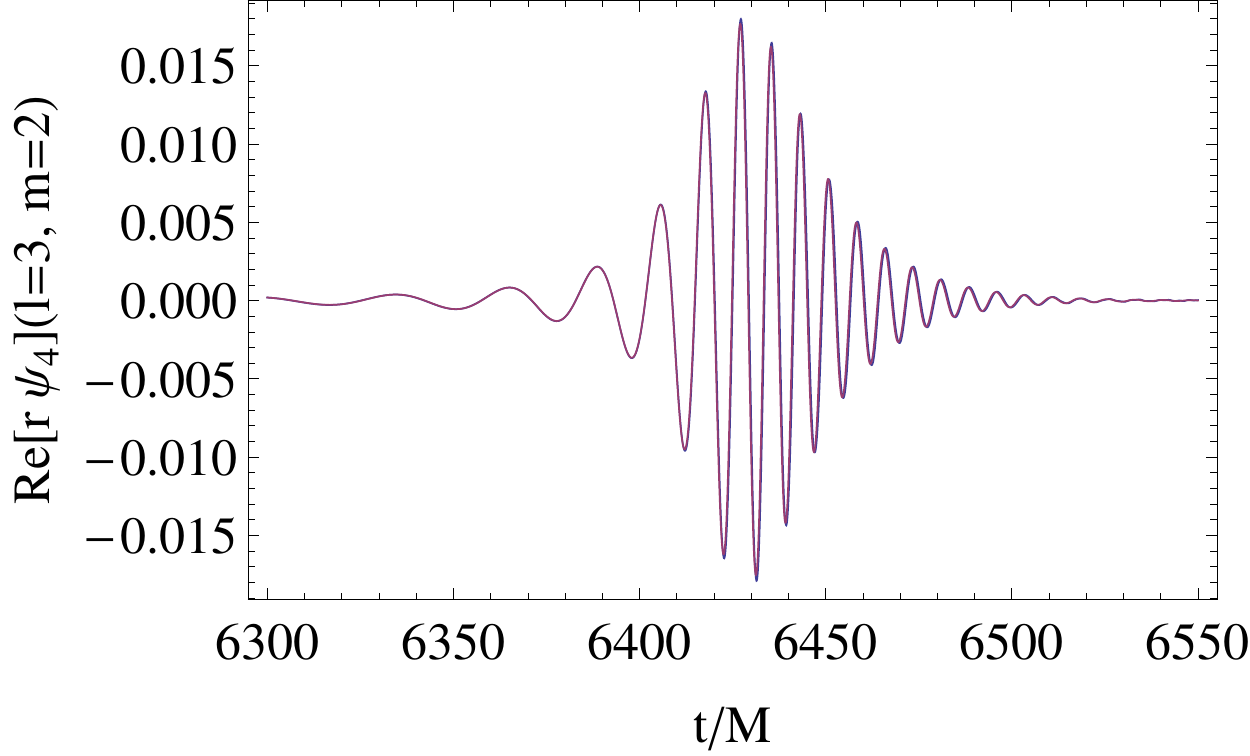}\\
  \caption{The new HiSpID simulation (blue) and the SXS simulation
    (red) of the $(\ell=3, m=2)$ mode of $\psi_4$ (real part). The HiSpID waveform was translated by
  $t\sim5000M$.}\label{fig:RIT_SXS_COMP_WAVE_32}
\end{figure}

The next largest modes after the $(\ell=2,m=\pm 2)$ modes are the
$(\ell=4, m=\pm4)$ modes. However, in our simulations, these show
significant effects of dissipation postmerger. We therefore compare
the $(\ell=3, m=2)$ mode instead.
As shown in
Fig.~\ref{fig:RIT_SXS_COMP_WAVE_32}, the agreement between \hispid and
SXS is quite good even for a higher-order mode. The overlap between
the $(\ell=3, m=2)$ modes is 0.998 [the constant $t_0$ was fixed by
maximizing the overlap of the $(\ell=2,m=2)$ modes, the maximum overlap of the
$(\ell=3, m=2)$ is 0.9998].

\subsection{Diagnostics}\label{sec:diagnostic}

\begin{figure}
  \includegraphics[width=0.9\columnwidth]{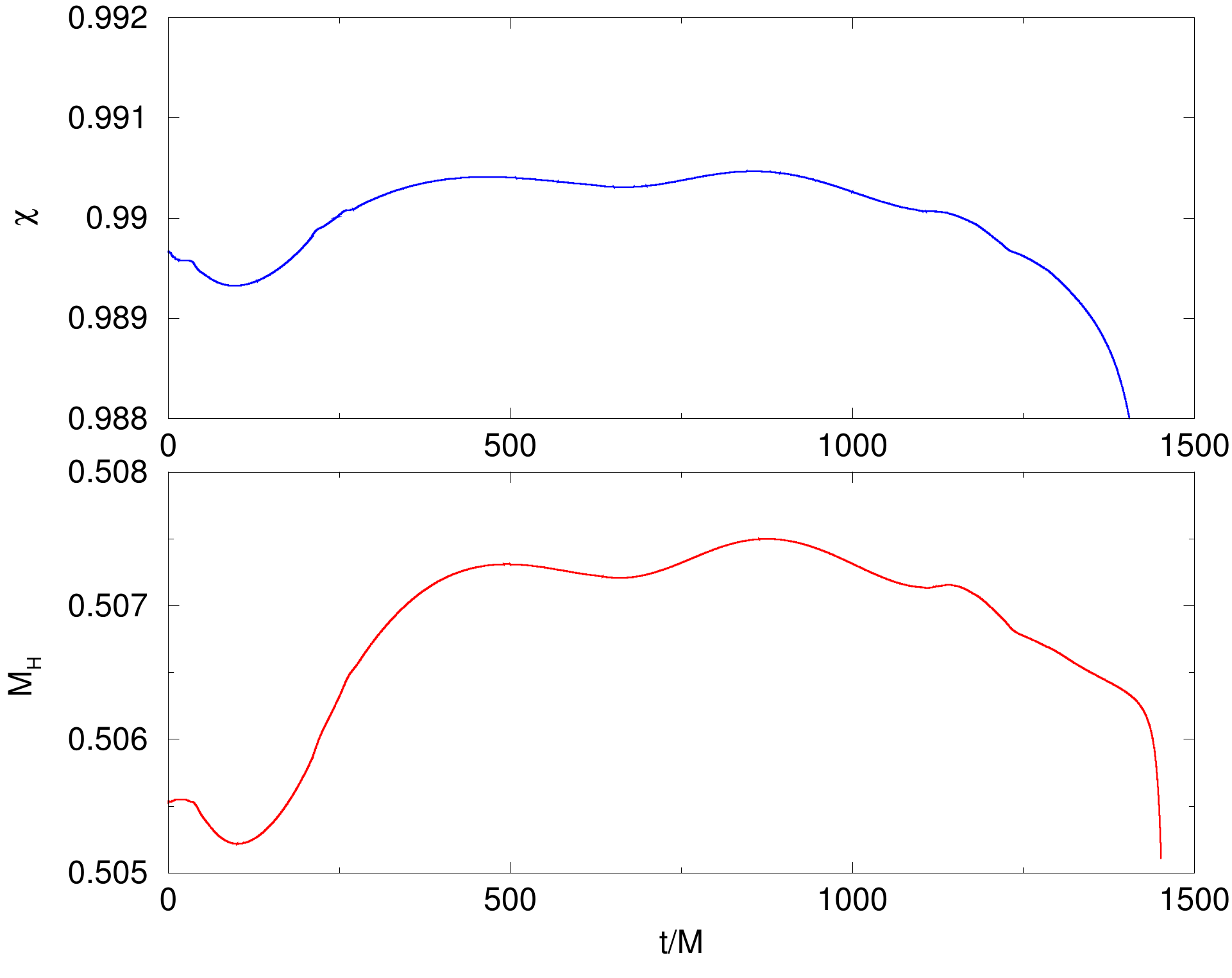}
  \caption{The dimensionless spin (top) and horizon (Christodoulou) mass 
  (bottom) for the two horizons in the binary.}\label{fig:hor}
\end{figure}

\begin{figure}
  \includegraphics[width=0.9\columnwidth]{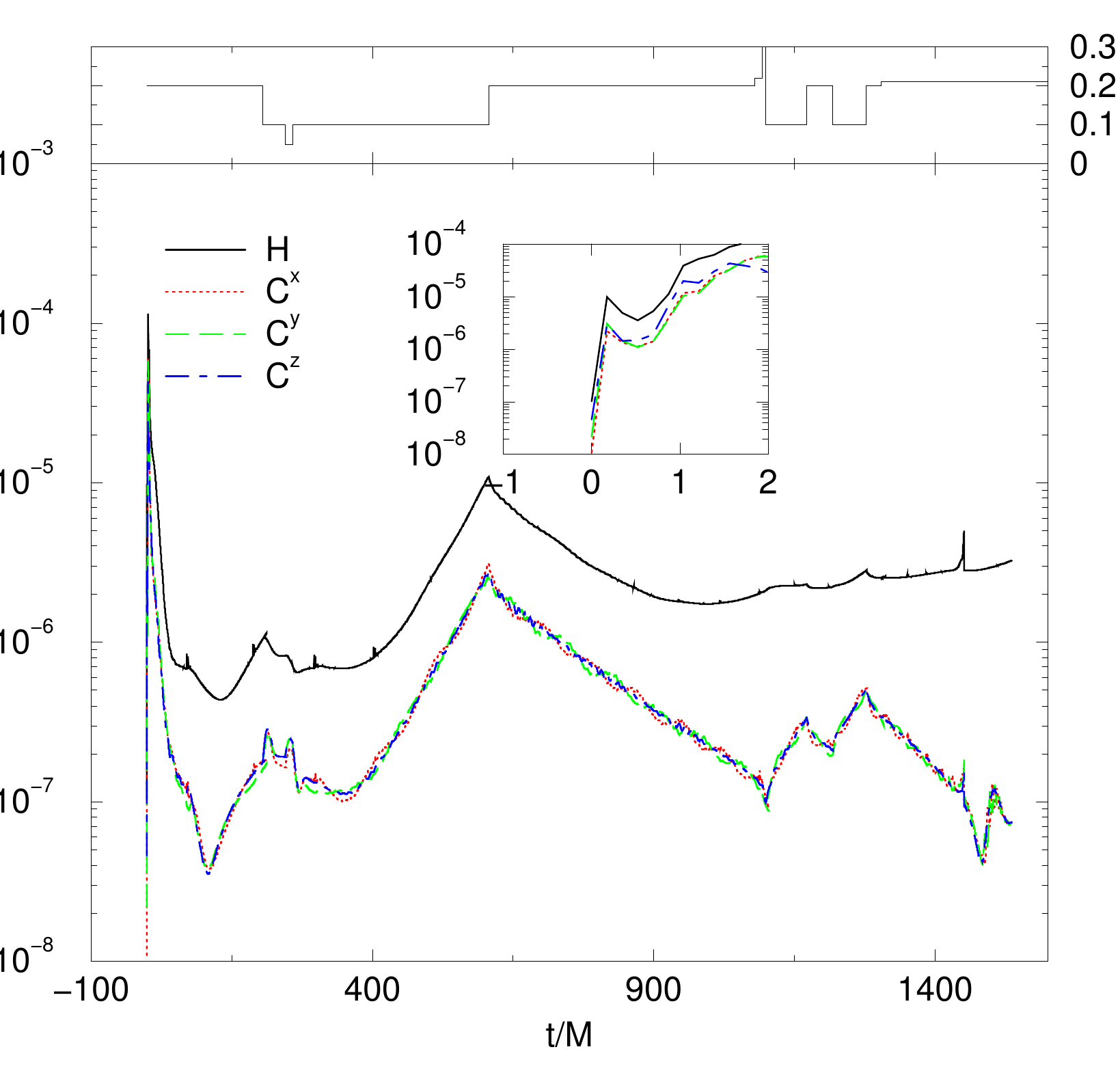}
  \caption{$L^2$ norm of the Hamiltonian and momentum constraints
    versus time. Note the rapid growth during the first 2M of
    evolution. The CCZ4 damping parameters $\kappa_{1,2}$ were
    adjusted during the evolution to suppress the constraint growths
  apparent at $t=400M - 600M$, and again at
$t=900M$. The top panel shows the value of $\kappa_1$ used during the
simulation.}\label{fig:const}
\end{figure}

One of the most important diagnostics for a BHB simulation is the
degree to which the constraints are satisfied and to what degree the
horizon masses and spins are conserved. In Fig.~\ref{fig:hor}, we show the individual horizon mass and
(dimensionless spin). Note that prior to merger, the spins are within
$\pm0.001$ of $0.99$ and the masses change by less than 0.2\%.
 In
Fig.~\ref{fig:const}, we show the $L^2$ norm of the Hamiltonian and
momentum constraints. Here the $L^2$ norm is over the region outside
the two horizons (or common horizon) and inside a sphere of radius
$30M$. Note how the constraints start small ($10^{-8}$) and quickly
increase to $10^{-4}$. This increase is due to unresolved features in
the initial data (i.e., the AMR grid cannot propagate high-frequency
data accurately). The constraints then damp, as is expected for CCZ4.
However, they start to exponentially blow up around $400M$. We found
that the parameters $\kappa_1$ and $\kappa_2$ had to be fine-tuned to
prevent this blow-up. We found that increasing the damping parameters
can effectively drive the constraints smaller for a short time, but
large values of $\kappa_i$ led to an exponential blowup of the
constraints at later times. We used a trial-and-error approach to
fine-tuning these parameters during the run. We show the values of 
$\kappa_1$ used during the evolution in the top of
Fig.~\ref{fig:const}.

\begin{figure}
  \includegraphics[width=0.9\columnwidth]{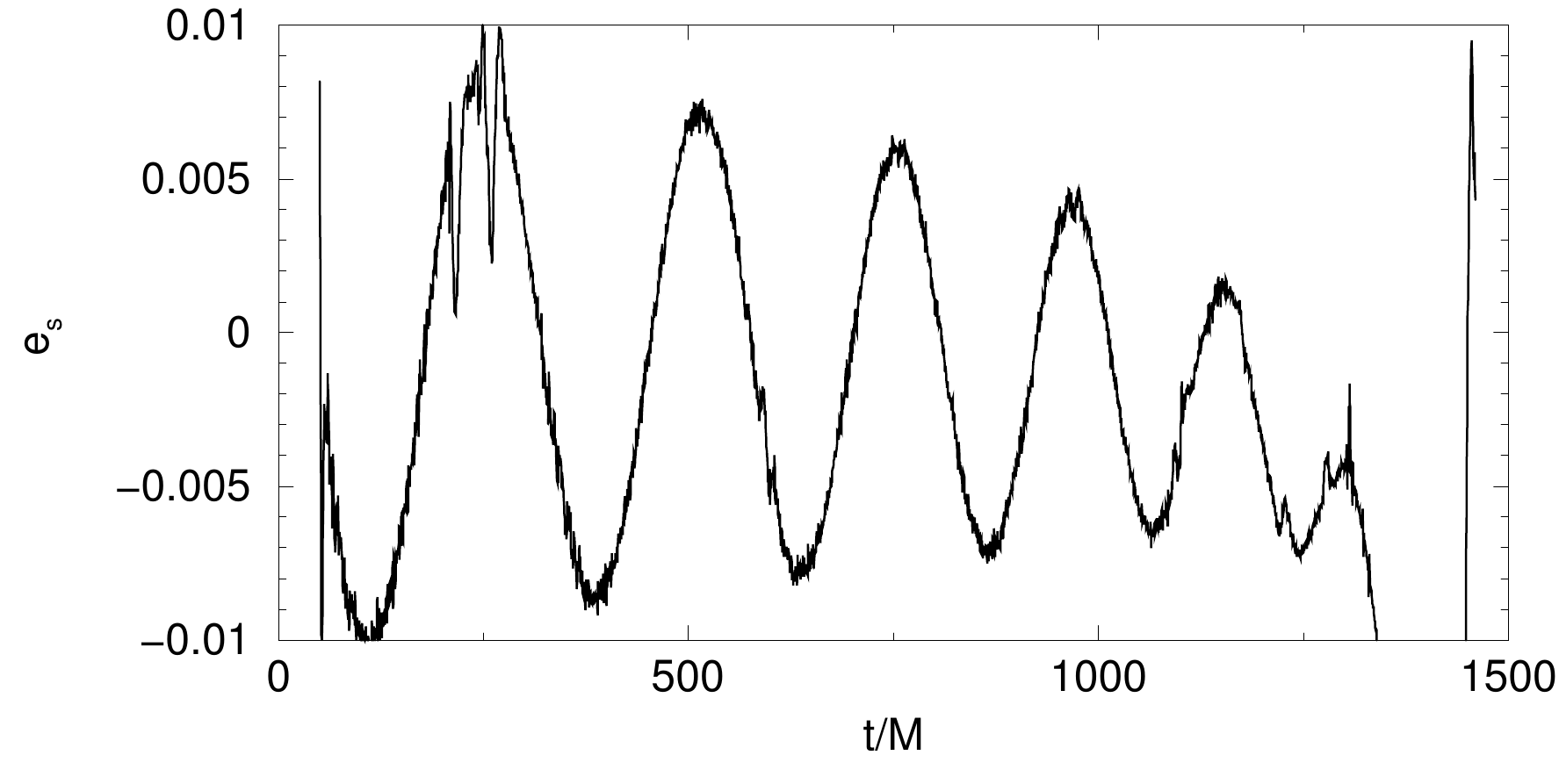}
  \caption{The eccentricity of the HiSpID UU99 simulation as measured
  using the approximation $e_s\approx s^2\ddot s$, where $s$ is the
proper distance of the part of the coordinate line segment connecting the centroids of the
two black holes that is outside both horizons. }\label{fig:ecc}
\end{figure}

One challenge with the \hispid data is obtaining low-eccentricity data
without performing an iterative procedure where the initial data are
evolved for a few orbits and then refined based on the measured
orbital evolution~\cite{Pfeiffer:2007yz,
  Buonanno:2010yk, Purrer:2012wy,
Buchman:2012dw}. In~\cite{Healy:2017zqj}, it was shown that relatively
low-eccentricity initial data parameters can be obtained using
higher-order post-Newtonian approximations. However, as shown in
Table~\ref{tab:id}, unlike for Bowen-York data, here we cannot specify
the initial momenta precisely. That is to say, the orbital angular
momentum of the background (i.e., prior to the inclusion of corrections
due to the fields $u$ and $b^i$) is significantly larger than the
final orbital angular momentum of the initial data. We compensate for
this by increasing the momentum parameters until the ADM angular
momentum matches the expected value based on quasicircular orbits.
However, we have no method of correcting for the radial momentum
(other than using an iterative evolution procedure).
Consequentially, the eccentricity of the initial data is relatively
high at $e\approx0.01$, as shown in Fig.~\ref{fig:ecc}. Of course, we
can run the data for a few orbits and then refine the parameters, but
such a procedure is computationally expensive. We are thus working on
improving the evolution efficiency.

\begin{figure}
  \includegraphics[width=0.9\columnwidth]{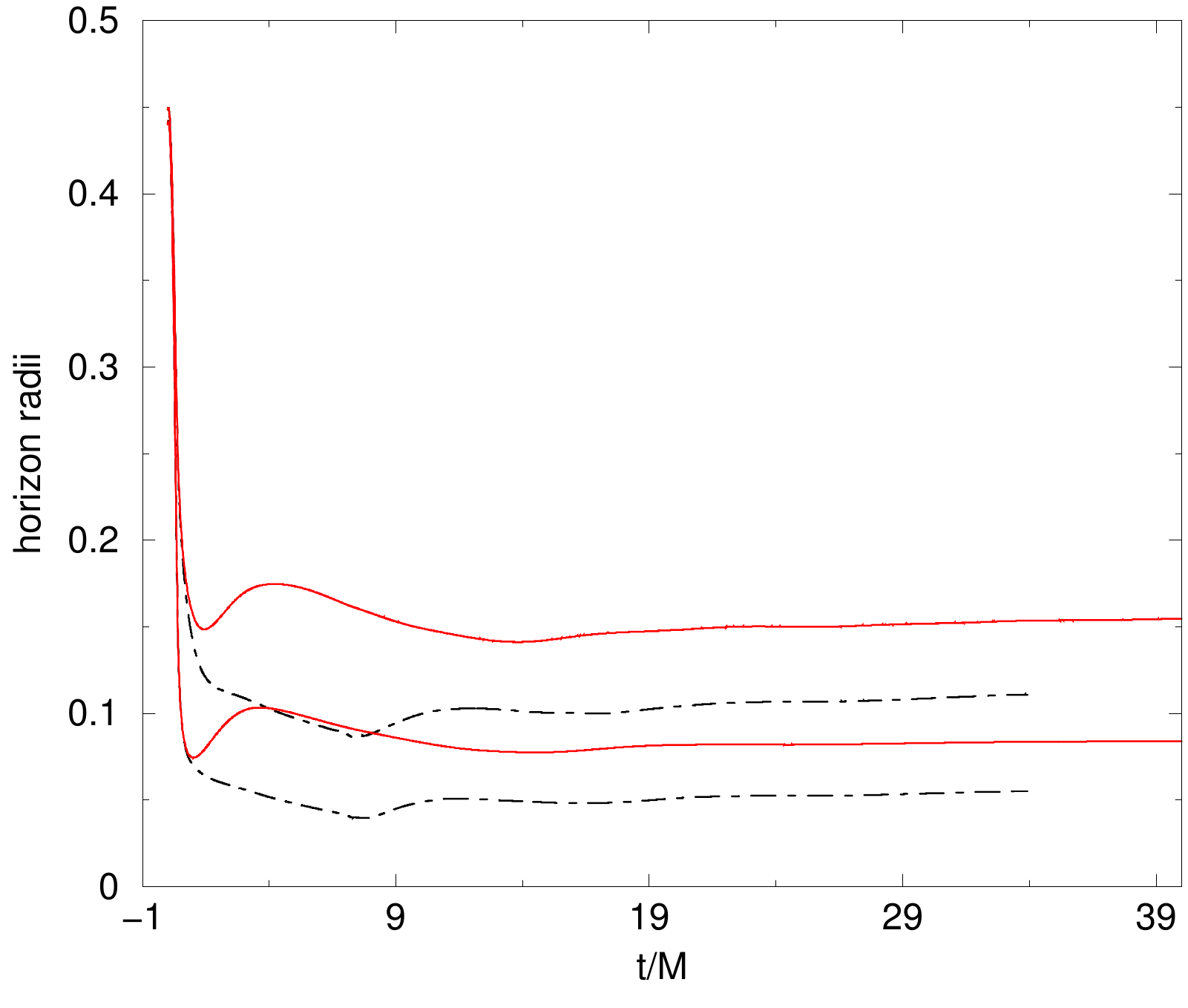}
  \caption{The coordinate radii (minimum and maximum) versus time for
    the standard 1+log lapse (dot-dashed curves) and the modified
    lapse condition used for the full simulation. Note that in both
    gauges there is an extremely rapid evolution of the horizon size
    and shape during the first few $M$ of evolution. The new gauge
  produces a horizon that is $\approx 8/5$ times larger.}
  \label{fig:horizon_radii}
\end{figure}

One method which we found was useful for increasing the run speed was
to change the lapse condition. When using the standard 1+log lapse,
the horizons are a factor of 0.625 as wide (see
Fig.~\ref{fig:horizon_radii}). Evolving the data with
horizons this small requires roughly a factor of 2 more in terms of
computational expense because an additional level of refinement is
needed. Using harmonic slicing leads to still larger horizons, but
this proved to be unstable. The rapid change in the gauge at early
time, as is evident in the size and shape of the horizon (see
Fig.~\ref{fig:horizon_radii}) may be responsible for the initial jump
in the constraint violations seen in Fig.~\ref{fig:const}.

\section{Discussion}\label{sec:discussion}

In this paper we demonstrated that it is possible to evolve black hole
binaries with nearly maximal spin using the ``moving puncture''
formalism. This means that comparative studies of these challenging
evolutions by the two main methods (the generalized harmonic approach
used by SXS and various flavors of the ``moving punctures'' approach
used by many other groups) to numerically solve the field equations of
general relativity field equations can now be performed.  Independent
comparison, along the lines explored in \cite{Lovelace:2016uwp}, have
been very successful in demonstrating the accuracy and correctness of
moderate-spin black hole simulations.  These new techniques also open
the possibility to explore a region of parameter space which is of
high interest for both astrophysical and gravitational wave studies.

\acknowledgments 
The authors thank M.Scheel for careful reading of the manuscript and
gratefully acknowledge the National Science Foundation (NSF) for financial support from Grants No.\
PHY-1607520, No.\ PHY-1707946, No.\ ACI-1550436, No.\ AST-1516150,
No.\ ACI-1516125.  This work used the Extreme Science and Engineering
Discovery Environment (XSEDE) [allocation TG-PHY060027N], which is
supported by NSF grant No. ACI-1548562.
Computational resources were also provided by the NewHorizons and
BlueSky Clusters at the Rochester Institute of Technology, which were
supported by NSF grants No.\ PHY-0722703, No.\ DMS-0820923, No.\
AST-1028087, and No.\ PHY-1229173.

\bibliographystyle{apsrev4-1}
\bibliography{../../../../Bibtex/references}

\end{document}